\documentstyle[12pt,fleqn]{article}

\catcode`\@=11
\@addtoreset{equation}{section}
\def\theequation{\arabic{section}.\arabic{equation}}
\newcommand{\beq}{\begin{equation}}
\newcommand{\eeq}{\end{equation}}
\newcommand{\beqy}{\begin{eqnarray}}
\newcommand{\eeqy}{\end{eqnarray}}

\def\tpi{\tilde{\pi}}
\def\tPi{\tilde{\Pi}}
\def \ut#1{\rlap{\lower1ex\hbox{$\sim$}}#1{}}
\def \UT#1{\rlap{\lower1ex\hbox{\scriptsize$\sim$}}#1{}}
\newcommand{\UI}[1]{^{\mbox{ } #1}}
\newcommand{\LI}[1]{_{\mbox{ } #1}}
\def\tN{\ut{N}}
\def\tM{\ut{M}}
\def\ep{\epsilon}
\def\otep{\tilde{\epsilon}}
\def\utep{\UT{\epsilon}}
\def\CA{{\cal A}}
\def\CB{{\cal B}}
\def\CF{{\cal F}}
\def\CD{{\cal D}}
\def\CH{{\cal H}}
\def\M3{M^{(3)}}
\def\Tr{{\rm Tr}}
\def\Str{{\rm STr}}
\begin{document}
\begin{flushright}
OU-HET/225\\hep-th/9511047\\November 1995
\end{flushright}
\vspace{0.5in}
\begin{center}\Large{\bf Ashtekar's formulation for $N=1,2$
supergravities\\
as \lq\lq constrained" BF theories}\\
\vspace{1cm}\renewcommand{\thefootnote}{\fnsymbol{footnote}}
\normalsize\ Kiyoshi Ezawa\footnote[1]
{Supported by JSPS.
e-mail address: ezawa@funpth.phys.sci.osaka-u.ac.jp}
        \setcounter{footnote}{0}
\vspace{0.5in}

        Department of Physics \\
        Osaka University, Toyonaka, Osaka 560, Japan\\
\vspace{0.1in}
\end{center}
\vspace{1.2in}
\baselineskip 17pt
\begin{abstract}

It is known that Ashtekar's formulation for pure Einstein gravity
can be cast into the form of a topological field theory,
namely the $SU(2)$ BF theory, with the
B-fields subject to an algebraic constraint. We extend this relation
between Ashtekar's formalism and BF theories to $N=1$ and $N=2$
supergravities. The relevant gauge groups in these cases become
graded Lie groups of $SU(2)$ which are generated by left-handed
local Lorentz transformations and left-supersymmetry transformations.
As a corollary of these relations, we provide topological solutions
for $N=2$ supergravity with a vanishing cosmological constant.
It is also shown that, due to the algebraic constraints, the
Kalb-Ramond symmetry which is characteristic of BF theories
breaks down to the symmetry under diffeomorphisms and
right-supersymmetry transformations.

\end{abstract}
\newpage

\baselineskip 20pt
\section{Introduction}

Since its birth in the mid eighties, Ashtekar's formulation
for canonical gravity\cite{ashte} has been vigorously investigated
by many people as a promising approach to
the nonperturbative quantum
gravity\cite{schil}. A merit of Ashtekar's formalism is that
the Hamiltonian constraint,
or the Wheeler-Dewitt equation\cite{dewitt},
takes a polynomial form in terms of new canonical variables.
Thus we expect that, using Ashtekar's formalism, we can solve the
constraint equations which has never been able to solve in the
conventional metric formulation\cite{ADM}.

In fact several types of solutions are found. They are roughly
classified into two types: \lq\lq loop solutions" which consist
of Wilson loops\cite{jacob}\cite{smol}\cite{ezawa} and;
\lq\lq topological solutions" which are also solutions for
a topological field theory\cite{bren}.
The latter type includes the Chern-Simons solution in the case
with a nonvanishing
cosmological constant $\Lambda$\cite{kodama}.
The existence of these
topological solutions suggests the relationship
between Ashtekar's formalism and a topological field theory,
namely the $SU(2)$ BF theory\cite{horo}. It was indeed shown that
Ashtekar's formalism can be obtained from the $SU(2)$
(strictly speaking the chiral $SL(2,{\bf C})$ ) BF theory
with the $B$ field subject to an algebraic constraint
\cite{capo}\cite{baez}\cite{ueno}.

Ashtekar's formalism is also applied to supergravities with
$N=1$\cite{jacob2}\cite{capo} and with $N=2$\cite{kuni}.
In the case of $N=1$ supergravity also, topological solutions
\cite{mats} including Chern-Simons solutions\cite{sano}\cite{shira}
were found. As for the $N=2$ case, only the Chern-Simons
solution was found\cite{sano}.
This fact implies the relation between $N=1,2$ supergravities and
BF theories with appropriate gauge groups. We expect that,
if these relations can be made transparent, we can make
further progress for investigation of quantum gravity
both technically and conceptually.

In this paper we show explicitly that Ashtekar's formulation
for $N=1$ and $N=2$ supergravities can indeed be cast into the
form of BF theories with the $B$-fields subject to algebraic
constraints. The relevant gauge groups in these cases are
provided by graded versions of $SU(2)$ which are
generated by left-handed local Lorentz transformations and
local left-supersymmetry transformations (plus $U(1)$ gauge
transformations in the $N=2$ case).
These relations not only elegantly explain the existence of
above mentioned topological solutions  but also predicts the
existence of topological solutions in the case with $N=2$ and
$\Lambda=0$.
We also show how the algebraic constraints for $B$-fields
breaks the Kalb-Ramond symmetry of the BF theories
\cite{KR}\cite{horo} down to the
symmetry under diffeomorphisms and right-supersymmetry
transformations.

The presentation of this paper is as follows.
Once the relations to the BF theories are established,
the arguments are almost parallel for the cases of pure gravity
and of $N=1,2$ supergravities. So only the $N=1$ case is dealt
with in detail. After briefly reviewing the relation between
pure gravity and the $SU(2)$ BF theory in \S 2, we derive the
action of $N=1$ chiral supergravity from that of the $GSU(2)$
BF theory in \S 3. Canonical quantization of the $GSU(2)$ BF theory
is also discussed in \S 3. $\Lambda=0$ case is a little peculiar
and the reduced phase space in this case is shown to be
the cotangent bundle over the moduli space
of flat $GSU(2)$ connections on the spatial manifold $\M3$
modulo gauge transformations.
In \S 4 we show that $N=2$ supergravity is obtained from
the BF theory whose gauge group being an appropriate
graded version of $SU(2)$, which we will henceforth call
$G^{2}SU(2)$. Unlike in the pure gravity and $N=1$ cases,
we cannot find the relation easily because the $N=2$ chiral
action involves a quadratic term in auxiliary fields.
Replacing this quadratic term by linear terms in auxiliary fields,
we make the relation to the $G^{2}SU(2)$ BF theory manifest.
For $N=2$ supergravity the $\Lambda=0$ case is somewhat different
from those in the $N=0,1$ cases because
the reduced phase space does not possess cotangent bundle structure.
We also provide the formal \lq\lq topological"
solutions for the $N=2$, $\Lambda=0$ case.
In \S 5 we discuss possibilities of future developments.

Finally we provide the convention used in this paper: i)
$\mu,\nu,\cdots$ stand for spacetime indices; ii) $a,b,\cdots$
are used for spatial indices; iii) $A,B,\cdots$ represent
left-handed $SL(2,{\bf C})$ spinor indices; iv) $i,j,\cdots$
denote indices for the  adjoint
representation of (the left-handed part of)
$SL(2,{\bf C})$; v) $\otep^{abc}$( $\utep_{abc}$) is the Levi-Civita
alternating tensor density of weight $+1$ ($-1$) with
$\otep^{123}=\utep_{123}=1$; vi) $\ep^{ijk}$
is the antisymmetric (pseudo-)tensor with $\ep^{123}=1$;
vii) $\ep^{AB}$ ($\ep_{AB}$) is the antisymmetric spinor with
$\ep^{12}=\ep_{12}=1$\footnote{These antisymmetric spinors are used
to raise and lower the spinor index:
$\varphi^{A}=\ep^{AB}\varphi_{B},
\quad \varphi_{A}=\varphi^{B}\ep_{BA}$.}; viii) relation between a
symmetric rank-2 spinor $\phi^{AB}$ and its equivalent vector
$\phi^{i}$ in the adjoint representation is given by
$\phi^{AB}=\phi^{i}(\frac{\sigma^{i}}{2i})^{AB}$, where
$(\sigma^{i})^{A}\LI{B}$ are the
Pauli matrices with $(\sigma^{i})^{A}\LI{C}(\sigma^{j})^{C}\LI{B}
=\delta^{ij}\delta^{A}_{B}+i\ep^{ijk}(\sigma^{k})^{A}\LI{B}$ ;
ix) $D=dx^{\mu}D_{\mu}$ denotes the covariant exterior derivative
with respect to the $SU(2)$ connection $A=A^{i}J_{i}$ and ;
x) indices located between $($ and $)$ ($[$ and $]$) are regarded
as symmetrized (antisymmetrized).

For simplicity we will restrict our analysis to the case where the
spacetime has the topology ${\bf R}\times\M3$ with $\M3$ being
a compact, oriented, 3 dimensional manifold without boundary.


\section{$SU(2)$ BF theory and Ashtekar's formalism}

In this section we provide a brief review of
the relationship between $SU(2)$
(or, strictly speaking, chiral $SL(2,{\bf C})$) BF theory and
Ashtekar's formulation for pure gravity \cite{capo} \cite{baez}
\cite{ueno}. We start with the action of the BF theory:
\beq
-iI_{BF}=\int\Tr(B\wedge F-\frac{\Lambda}{6}B\wedge B),
\label{eq:bfac}
\eeq
where $B=\Sigma^{i}J_{i}=\frac{1}{2}\Sigma^{i}_{\mu\nu}dx^{\mu}\wedge
dx^{\nu}J_{i}$ is an $SU(2)$ Lie algebra-valued two-form and
$F=dA+A\wedge A$ is the curvature two-form of an $SU(2)$ connection
$A=A^{i}J_{i}=A^{i}_{\mu}dx^{\mu}J_{i}$\footnote{
$J_{i}$ denote the $SU(2)$ generators subject to the commutation
relations $[J_{i},J_{j}]=\ep_{ijk}J_{k}$. $\Tr$ is the
$SU(2)$-invariant bilinear form: $\Tr(J_{i}J_{j})=\delta_{ij}$}.

This action is invariant under $SU(2)$ gauge (or
left-handed local Lorentz) transformations
\beqy
\delta_{\theta}B&=&[\theta,B]\nonumber \\
\delta_{\theta}A&=&-D\theta\equiv-d\theta-[A,\theta],
\eeqy
where $\theta=\theta^{i}J_{i}$ is an $SU(2)$-valued scalar. Action
(\ref{eq:bfac}) has the additional symmetry, the (generalized)
Kalb-Ramond symmetry\cite{KR}\cite{horo}:
\beqy
\delta_{\phi}B&=&-D\phi\equiv
-d\phi-A\wedge\phi-\phi\wedge A\nonumber \\
\delta_{\phi}A&=&-\frac{\Lambda}{3}\phi,\label{eq:KR}
\eeqy
where $\phi=\phi^{i}J_{i}$ is an $SU(2)$-valued one form.
This Kalb-Ramond symmetry includes the symmetry under the
diffeomorphisms. We use as $\phi$ the $B$-field-dependent parameter
$\phi_{\mu}=v^{\nu}B_{\mu\nu}$. Then by using equations of motion
$$
F-\frac{\Lambda}{3}B=DB=0,
$$
we obtain the infinitesimal diffeomorphism generated by the vector
$v=v^{\mu}\frac{\partial}{\partial x^{\mu}}$ plus the $SU(2)$ gauge
transformation generated by $\theta=v^{\mu}A_{\mu}$:
\beqy
\delta_{\phi}B|_{\phi_{\mu}=v^{\nu}B_{\mu\nu}}&=&
{\cal L}_{v}B+\delta_{\theta}B|_{\theta=v^{\mu}A_{\mu}}\nonumber \\
\delta_{\phi}A|_{\phi_{\mu}=v^{\nu}B_{\mu\nu}}&=&
{\cal L}_{v}A+\delta_{\theta}A|_{\theta=v^{\mu}A_{\mu}},
\eeqy
where ${\cal L}_{v}$ denotes the Lie derivative with respect to the
vector $v$. The derivation of these equations
needs the equations of motion. However, the equations of
motion for the BF theory are either first class constraints or
equations which yield conditions for the temporal
components of the fields. As is seen shortly,
in the canonical formalism, temporal components are
considered to be Lagrange multipliers which play the role of
the gauge parameters.
The diffeomorphism invariance is thus considered to be a particular
form of the Kalb-Ramond symmetry as far as the physical contents of
the BF theory are concerned.

In order to rewrite the action (\ref{eq:bfac}) in the canonical form,
we simply identify the zeroth coordinate $x^{0}$ with time $t$.
The result is
\beqy
-iI_{BF}&=&\int dt\int_{\M3}d^{3}x
\Tr[\tpi^{a}\dot{A}_{a}+A_{t}G+\Sigma_{ta}\Phi^{a}]
\nonumber \\
&=&\int dt\int_{\M3}d^{3}x(\tpi^{ai}\dot{A}^{i}_{a}+A^{i}_{t}G^{i}
+\Sigma_{ta}^{i}\Phi^{ai}),\label{eq:bfcano}
\eeqy
where we have set $\tpi^{a}=\tpi^{ai}J_{i}
\equiv\frac{1}{2}\otep^{abc}B_{bc}$ and $\dot{A}=
\frac{\partial}{\partial t}A$. This system involves two types of
first class constraints. Gauss' law constraint
\beq
G=G^{i}J_{i}\equiv D_{a}\tpi^{a}
\eeq
generates the
$SU(2)$ gauge transformations and the remaining constraint
\beq
\Phi^{a}=\Phi^{a}_{i}J_{i}\equiv \frac{1}{2}\otep^{abc}F_{bc}-
\frac{\Lambda}{3}\tpi^{a}
\eeq
generates the Kalb-Ramond transformations.

Let us now quantize this system following Dirac's quantization
procedure\cite{dirac}.
We first read off canonical commutation relations
from the symplectic structure. The result is
$$
[\hat{A}^{i}_{a}(x),\hat{\tpi}^{bj}(y)]=\delta^{b}_{a}\delta^{ij}
\delta^{3}(x,y).
$$
If we use as wavefunctions the functionals
of the connection $A^{i}_{a}$,
the conjugate momenta are represented by functional differentiations:
\beq
\hat{\tpi}^{a}_{i}(x)\cdot\Psi[A^{i}_{a}]=
-\frac{\delta}{\delta A_{a}^{i}(x)}\Psi[A^{i}_{a}].
\eeq

Next we impose the first class constraints as conditions to which
the physical wavefunctions are subject.
Gauss' law constraint simply tells
us that the wavefunctions be $SU(2)$ gauge invariant. The other
constraint can also be solved easily.

For $\Lambda=0$ this
is solved by the wavefunctions with support only on the flat
connections. The formal solutions for $\Lambda=0$ case are given by
\beq
\Psi[A]=\psi[A_{a}^{i}]\prod_{x\in\M3}\prod_{i,a}
\delta(\otep^{abc}F^{i}_{bc}(x)),
\eeq
where $\psi$ is an arbitrary $SU(2)$ gauge invariant functional
of the connection. This solution coincides with that obtained
in ref.\cite{bren}.
This is effectively equivalent to dealing with
the functions on the moduli space of flat $SU(2)$ connections
modulo (the identity-connected component of)
gauge transformations\footnote{
Because the constraints are at most linear in the conjugate momenta
$\tpi^{a}$, we expect that Dirac's quantization yields the same
result as the reduced phase space quantization, up to minor
subtleties\cite{ashte2}. Particularly in the $\Lambda=0$ case,
the result of these two quantizations should be
identical because the reduced phase space turns out to be the
cotangent bundle on the moduli space of flat $SU(2)$ connections.}.

For $\Lambda\neq0$ this remaining constraint has a unique solution
\beq
\Psi[A]=\exp[-\frac{3}{2\Lambda}\int_{\M3}\Tr(AdA+
\frac{2}{3}A\wedge A\wedge A)],
\eeq
which coincides with the Chern-Simons solution found in
ref.\cite{kodama}.

Ashtekar's formulation for pure gravity is obtained from the action
(\ref{eq:bfac}), accompanied by the following algebraic constraint
on the $B$-field (we set $\Sigma^{AB}=\Sigma^{i}
(\frac{\sigma^{i}}{2i})^{AB}$):
\beq
\Sigma^{(AB}\wedge\Sigma^{CD)}=0.
\eeq
Solving this algebraic constraint for $\Sigma^{i}_{ta}$ and
substituting the result into the action(\ref{eq:bfcano}), we find
\beq
-iI_{Ash}=\int dt\int_{\M3}d^{3}x\Tr\left(\begin{array}{r}
\tpi^{a}\dot{A}_{a}+A_{t}D_{a}\tpi^{a}
-i\tN\frac{1}{2}\tpi^{b}\tpi^{c}(F_{bc}-
\frac{\Lambda}{3}\utep_{bca}\tpi^{a})\\
+N^{b}\tpi^{c}(F_{bc}-
\frac{\Lambda}{3}\utep_{bca}\tpi^{a})\end{array}\right).
\eeq
This is nothing but the action for  Ashtekar's formalism.
Thus we easily see that the solutions to the $SU(2)$ BF theory
are necessarily included in the solution space of
Ashtekar's constraints provided that we take the ordering
with the momenta $\tpi^{ai}$ to the left. This seems to be natural
because we know that the constraint algebra formally closes
under such ordering.


\section{$GSU(2)$ BF theory and Ashtekar's formulation
for $N=1$ supergravity}

In this section we show explicitly that $N=1$ supergravity in
Ashtekar's form can be cast into the form of the $GSU(2)$ BF theory
with the $B$ field subject to algebraic constraints.

\subsection{$GSU(2)$ BF theory}

We start with the following BF action
\beq
-iI^{N=1}_{BF}=\int\Str(\CB\wedge\CF-\frac{g^{2}}{6}\CB\wedge\CB),
\label{eq:BFac}
\eeq
where $\CB=\Sigma^{i}J_{i}-\frac{1}{\lambda g}\chi^{A}J_{A}$
is a $GSU(2)$-valued two-form and $\CF=d\CA+\CA\wedge\CA$ is
the curvature two form of the $GSU(2)$ connection $\CA=A^{i}J_{i}+
\psi^{A}J_{A}$.\footnote{Note that $\chi^{A}$ and $\psi^{A}$ are
Grassmann odd fields. Whether an object is Grassmann even or odd
can be determined by whether the number of its Lorentz
spinor indices is even or odd.} $(J_{i},J_{A})$ are the generators
of the graded Lie algebra $GSU(2)$\cite{pais}:
\beq
[J_{i},J_{j}]=\ep_{ijk}J_{k},\quad
[J_{i},J_{A}]=(\frac{\sigma^{i}}{2i})_{A}\UI{B}J_{B},\quad
\{J_{A},J_{B}\}=-2\lambda g(\frac{\sigma^{i}}{2i})_{AB}J_{i},
\label{eq:GSU2}
\eeq
where $\{\mbox{ , } \}$ denotes the anti-commutation relation.
$\Str$ stands for the $GSU(2)$ invariant bilinear form
which is unique up to an overall constant factor
\beqy
&&\Str(J_{i}J_{j})=\delta_{ij},\quad
\Str(J_{A}J_{B})=-2\lambda g\ep_{AB},\nonumber \\
&&\Str(J_{A}J_{i})=\Str(J_{i}J_{A})=0.\label{eq:bin}
\eeqy
If we use eqs.(\ref{eq:GSU2}) and (\ref{eq:bin}) and
rewrite the action(\ref{eq:BFac}) in terms of component fields,
the result is as follows
\beq
-iI^{N=1}_{BF}=\int\left(\begin{array}{r}
\Sigma^{i}\wedge(F^{i}+\lambda g(\frac{\sigma^{i}}{2i})_{AB}
\psi^{A}\wedge\psi^{B})+2\chi^{A}\wedge D\psi_{A}\\
-\frac{g^{2}}{6}\Sigma^{i}\wedge\Sigma^{i}
-\frac{g}{3\lambda}\chi_{A}\wedge\chi^{A}
\end{array}\right).\label{eq:ASH1}
\eeq

This action (\ref{eq:BFac}), or equivalently
the action (\ref{eq:ASH1}),
necessarily possesses the symmetry under $GSU(2)$ gauge
transformations
\beqy
\delta_{\rho}\CA&=&-\CD\rho\equiv-d\rho-[\CA,\rho],\nonumber \\
\delta_{\rho}\CB&=&[\rho,\CB],\label{eq:GAUGE}
\eeqy
where $\rho=\theta^{i}J_{i}+\ep^{A}J_{A}$ is a $GSU(2)$-valued
scalar, and the Kalb-Ramond symmetry
\beqy
\delta_{\xi}\CA&=&-\frac{g^{2}}{3}\xi,\nonumber \\
\delta_{\xi}\CB&=&-\CD\xi\equiv-d\xi-\CA\wedge\xi-\xi\wedge\CA,
\label{eq:KR1}
\eeqy
where $\xi=\phi^{i}J_{i}-\frac{1}{\lambda g}\eta^{A}J_{A}$ is
a $GSU(2)$-valued one-form. Of course these transformations can be
translated in terms of component fields. The $GSU(2)$ gauge
transformation (\ref{eq:gauge}) is
\beqy
\delta_{\rho}A^{i}&=&-D\theta^{i}+
2\lambda g(\frac{\sigma^{i}}{2i})_{AB}\ep^{A}\psi^{B}\nonumber \\
\delta_{\rho}\psi^{A}&=&\theta^{i}(\frac{\sigma^{i}}{2i})^{A}\LI{B}
\psi^{B}-D\ep^{A}\nonumber \\
\delta_{\rho}\Sigma^{i}&=&\ep^{ijk}\theta^{j}\Sigma^{k}-2
(\frac{\sigma^{i}}{2i})_{AB}\ep^{A}\chi^{B}\nonumber \\
\delta_{\rho}\chi^{A}&=&\theta^{i}(\frac{\sigma^{i}}{2i})^{A}\LI{B}
\chi^{B}+\lambda g\ep^{B}
(\frac{\sigma^{i}}{2i})_{B}\UI{A}\Sigma^{i}.\label{eq:gauge}
\eeqy
And the Kalb-Ramond symmetry is decomposed as
\beqy
\delta_{\xi}A^{i}&=&-\frac{g^{2}}{3}\phi^{i} \nonumber \\
\delta_{\xi}\psi^{A}&=&\frac{g}{3\lambda}\eta^{A} \nonumber \\
\delta_{\xi}\Sigma^{i}&=&-D\phi^{i}+2(\frac{\sigma^{i}}{2i})_{AB}
\eta^{A}\wedge\psi^{B} \nonumber \\
\delta_{\xi}\chi^{A}&=&\lambda g\phi^{i}
(\frac{\sigma^{i}}{2i})^{A}\LI{B}\wedge\psi^{B}-D\eta^{A}.
\label{eq:kr1}
\eeqy
In almost the same way as in the $SU(2)$ case, we can show that a
diffeomorphism is generated by the Kalb-Ramond
transformation (\ref{eq:KR1})
(or the transformation (\ref{eq:kr1})) with
$\xi_{\mu}=v^{\nu}\CB_{\mu\nu}$, up to a $GSU(2)$
gauge transformation generated by $\rho=v^{\mu}\CA_{\mu}$.

Let us now investigate the canonical formalism. Similarly to the
$SU(2)$ case we perform (3+1)-decomposition of the action.
The result is
\beqy
-iI^{N=1}_{BF}&\!\!\!=\!\!\!&
\int dt\int_{\M3}\Str(\tPi^{a}\dot{\CA}_{a}+\CA_{t}
{\bf G}+\CB_{ta}{\bf \Phi}^{a})\nonumber \\
&\!\!\!=\!\!\!&\int dt\int_{\M3}
(\tpi^{ai}\dot{A}^{i}_{a}+2\tpi^{A}\dot{\psi}_{aA}+A_{t}^{i}G^{i}
-2\psi_{tA}L^{A}+\Sigma_{ta}^{i}\Phi^{ai}-2\chi_{taA}\Phi^{aA}),
\label{eq:caac1}
\eeqy
where we have set $\tPi^{a}=\frac{1}{2}\otep^{abc}\CB_{bc}=
\tpi^{ai}J_{i}-\frac{1}{\lambda g}\tpi^{aA}J_{A}$. From the
symplectic potential
\beq
\Theta=i\int_{\M3}d^{3}x\Str(\tPi^{a}\delta\CA_{a})
=i\int_{\M3}d^{3}x(\tpi^{ai}\delta A^{i}_{a}
+2\tpi^{aA}\delta\psi_{aA})\label{eq:sympl}
\eeq
we can read off Poisson brackets between the canonical variables:
\beq
\{A^{i}_{a}(x),\tpi^{bj}(y)\}_{PB}=
-i\delta^{b}_{a}\delta^{ij}\delta^{3}(x,y),\quad
\{\psi_{aA}(x),\tpi^{bB}(y)\}_{PB}=
\frac{-i}{2}\delta^{b}_{a}\delta^{B}_{A}\delta^{3}(x,y),
\label{eq:PB1}
\eeq
with the rest being zero.

There are two types of first class constraints. One is the
Gauss' law constraint which generates $GSU(2)$ gauge
transformations of the canonical variables
under the Poisson bracket
\beqy
{\bf G}&=&\CD_{a}\tPi^{a}=G^{i}J_{i}-\frac{1}{\lambda g}L^{A}J_{A},
\nonumber \\
G^{i}&=&D_{a}\tpi^{ai}-2(\frac{\sigma^{i}}{2i})_{AB}
\psi^{A}_{a}\tpi^{aB}\nonumber \\
L^{A}&=&D_{a}\tpi^{aA}+\lambda g\tpi^{ai}
(\frac{\sigma^{i}}{2i})^{A}\LI{B}\psi^{B}_{a}.\label{eq:gauss1}
\eeqy
And the other is the constraint which
generates the Kalb-Ramond transformations
\beqy
{\bf \Phi}^{a}&=&\frac{1}{2}\otep^{abc}\CF_{bc}
-\frac{g^{2}}{3}\tPi^{a}\nonumber \\
&=&\Phi^{ai}J_{i}+\Phi^{aA}J_{A},\nonumber \\
\Phi^{ai}&=&\frac{1}{2}\otep^{abc}(F^{i}_{bc}+2\lambda g
(\frac{\sigma^{i}}{2i})_{AB}\psi^{A}_{b}\psi^{B}_{c})
-\frac{g^{2}}{3}\tpi^{ai}\nonumber \\
\Phi^{aA}&=&\otep^{abc}D_{b}\psi_{c}^{A}+
\frac{g}{3\lambda}\tpi^{aA}.\label{eq:ckr1}
\eeqy
These constraints indeed form a closed algebra under the
Poisson bracket. To see this it is convenient to use smeared
constraints
\beqy
{\bf G}(\rho)&\equiv&i\int_{\M3}d^{3}x\Str(\rho{\bf G})
=i\int_{\M3}d^{3}x(\theta^{i}G^{i}-2\ep_{A}L^{A})\nonumber \\
&\equiv&G^{i}(\theta^{i})+L^{A}(\ep_{A}), \label{eq:SGAUSS}\\
{\bf \Phi}(\xi)&\equiv&i\int_{\M3}d^{3}x\Str(\xi_{a}{\bf \Phi})
=i\int_{\M3}d^{3}x(\phi_{a}^{i}\Phi^{ai}-2\eta_{aA}\Phi^{aA})
\nonumber \\
&\equiv&\Phi^{ai}(\phi^{i}_{a})+\Phi^{aA}(\eta_{aA}).
\label{eq:SKR}
\eeqy
The calculation of Poisson brackets is not so difficult
if we recall that these smeared constraints
generate the $GSU(2)$ gauge transformation (\ref{eq:GAUGE}) and
the Kalb-Ramond transformation (\ref{eq:KR1}) on
the canonical variables. The result is neatly written as
\beqy
\{{\bf G}(\rho),{\bf G}(\rho^{\prime})\}_{PB}&=&
{\bf G}([\rho,\rho^{\prime}]),\nonumber \\
\{{\bf\Phi}(\xi),{\bf G}(\rho)\}_{PB}&=&
{\bf\Phi}([\xi,\rho]),\nonumber \\
\{{\bf \Phi}(\xi),{\bf\Phi}(\xi^{\prime})\}_{PB}&=&0.
\label{eq:conal1}
\eeqy
Of course these involve the whole information on the constraint
algebra written in terms of component fields. For instance
the Poisson algebra between the components of the
Gauss' law constraint reads as
\beqy
\{iG^{i}(x),iG^{j}(y)\}_{PB}&=&\delta^{3}(x,y)\ep^{ijk}
(iG^{k}(x)) \nonumber \\
\{iG^{i}(x),-2iL_{A}(y)\}_{PB}&=&\delta^{3}(x,y)
(\frac{\sigma^{i}}{2i})_{A}\UI{B}(-2iL_{B}(x))\nonumber \\
\{-2iL_{A}(x),-2iL_{B}(y)\}_{PB}&=&\delta^{3}(x,y)(-2\lambda g)
(\frac{\sigma^{i}}{2i})_{AB}(iG^{i}(x)).\label{eq:const1}
\eeqy
This precisely coincides with the $GSU(2)$ algebra.

One may suspect that the case with $g=0$ need a careful
consideration because the definition of $\CB$ is singular
at $g=0$. However, we do not have to be so nervous since
no negative power of $g$ appears either in the action
or in the smeared constraints, provided that they are
expressed in terms of component fields. Indeed if
we start with the action for the component fields
\beq
-iI^{N=2}_{BF}=\int(\Sigma^{i}\wedge F^{i}
+2\chi^{a}\wedge D\psi_{A}),
\eeq
and consider the linear combinations of the constraints
appearing in the last expressions of
eqs.(\ref{eq:SGAUSS}) and (\ref{eq:SKR}),
the result of the
constraint algebra reproduces the $g\rightarrow0$
limit of the equations(\ref{eq:conal1})(\ref{eq:const1}).

One of the properties characteristic of the $g=0$ case is that
the symplectic potential (\ref{eq:sympl}) is inherited to the
reduced phase space\footnote{Reduced phase space is
the quotient space of the constraint surface modulo
gauge transformations in a broader sense. The constraint surface is
the subspace of the phase space on which the first order
constraints vanish. Gauge transformations in a broader sense are
the transformations generated by the first order constraints.}.
To see this explicitly we compute the transformation property
of the symplectic potential under the $GSU(2)$ gauge transformations
and the Kalb-Ramond transformations. We find
\beqy
\delta_{\theta}\Theta&=&i\int_{\M3}d^{3}xG^{i}\delta\theta^{i},
\nonumber \\
\delta_{\ep}\Theta&=&2i\int_{\M3}d^{3}xL^{A}\delta\ep_{A},
\nonumber \\
\delta_{\phi}\Theta&=&i\int_{M3}d^{3}x\phi^{i}_{a}\delta\Phi^{ai},
\nonumber \\
\delta_{\eta}\Theta&=&
2i\int_{\M3}d^{3}x\eta^{A}_{a}\delta\Phi_{A}^{a}.
\eeqy
These expressions vanish on the constraint surface $G^{i}=L^{A}=
\Phi^{ai}=\Phi^{a}_{A}=0$. This implies that the reduced phase space
has a well-defined cotangent bundle structure. The base space of
this cotangent bundle is provided by the reduced configuration space,
which in this case turns out to be the moduli space ${\cal N}_{0}$
of flat $GSU(2)$
connections $\CA_{a}=A^{i}_{a}J_{i}+\psi^{A}_{a}J_{A}$ on $\M3$
modulo $GSU(2)$ gauge transformations. The reduced phase space in the
$g=0$ case is therefore the cotangent bundle $T^{\ast}{\cal N}_{0}$
over the moduli space ${\cal N}_{0}$ of flat $GSU(2)$ connections.

To quantize this system canonically, we have only to replace
($i$-times of )the basic poisson brackets (\ref{eq:PB1}) by the
commutation relations. If we use as wave functions the functionals
$\Psi[\CA_{a}]$ of the connection $\CA_{a}=A^{i}_{a}J_{i}
+\psi^{A}_{a}J_{A}$, the conjugate momenta become the functional
derivatives:
\beq
\hat{\tpi^{ai}}(x)\cdot\Psi[\CA]=
-\frac{\delta}{\delta A_{a}^{i}(x)}\Psi[\CA],\quad
\hat{\tpi^{aA}}(x)\cdot\Psi[\CA]=
\frac{1}{2}\frac{\delta}{\delta\psi_{aA}(x)}\Psi[\CA].
\eeq

Next we solve the constraint equations.
Gauss' law constraint
\beq
\hat{G}^{i}\cdot\Psi[\CA]=\hat{L}^{A}\cdot\Psi[\CA]=0
\eeq
requires the wavefunctions to be invariant under the
(identity-connected component of the) $GSU(2)$ gauge transformations.
The remaining constraint
\beq
\hat{\Phi}^{ai}\cdot\Psi[\CA]=\hat{\Phi}^{aA}\cdot\Psi[\CA]=0
\label{eq:kreq}
\eeq
can easily be solved (at least formally).

For $g=0$, this constraint
requires the wavefunctions to have support only on the flat
$GSU(2)$ connections. The solutions to all the constraints are
therefore provided formally by
\beq
\Psi[\CA]=F[\CA_{a}]\prod_{x\in\M3}
\left(\prod_{a,i}\delta(\otep^{abc}F^{i}_{bc}(x))
\prod_{a,A}\delta(\otep^{abc}D_{b}\psi^{A}_{c}(x))\right),
\label{eq:topo1}
\eeq
where $F[\CA_{a}]$ is an arbitrary $GSU(2)$-invariant functional
of the connection $\CA_{a}$. Owing to the delta functions
$F[\CA]$ reduces to the function on the moduli space
${\cal N}_{0}$ of flat $GSU(2)$ connections. Thus, naively,
these solutions are considered to be \lq\lq Fourier transforms" of
the topological solutions found in ref.\cite{mats}.

For $g\neq 0$, we can rewrite the equation (\ref{eq:kreq}) as
\beqy
\left(\frac{\delta}{\delta A_{a}^{i}}+\frac{3}{2g^{2}}
\Str(J_{i}\otep^{abc}\CF_{bc})\right)\cdot\Psi[\CA_{a}]&=&0,
\nonumber \\
\left(\frac{\delta}{\delta\psi_{aA}}-\frac{3}{2g^{2}}
\Str(J^{A}\otep^{abc}\CF_{bc})\right)\cdot\Psi[\CA_{a}]&=&0.
\eeqy
These equations have a unique solution, which is
\beq
\Psi[\CA]=e^{-\frac{3}{2g^{2}}W_{CS}^{N=1}},\label{eq:CS1}
\eeq
where $W_{CS}^{N=1}$ is the Chern-Simons functional for the
$GSU(2)$ connection $\CA$:
\beqy
W_{CS}^{N=1}&=&\int_{\M3}\Str(\CA d\CA+
\frac{2}{3}\CA\wedge\CA\wedge\CA)\nonumber \\
&=&\int_{\M3}(A^{i}dA^{i}+\frac{1}{3}\ep^{ijk}A^{i}\wedge A^{j}
\wedge A^{k}-2\lambda g\psi^{A}\wedge D\psi_{A}).
\eeqy
The solution (\ref{eq:CS1}) coincides with the $N=1$ Chern-Simons
solution found in refs.\cite{sano}\cite{shira}.


\subsection{Ashtekar's formalism for $N=1$ supergravity}

We are now in a position to discuss the relation of the $N=1$
Ashtekar formalism to the $GSU(2)$ BF theory. First we notice that
the action (\ref{eq:ASH1}) is identical to the chiral action
for N=1 supergravity \cite{capo} with a cosmological constant
$\Lambda=g^{2}$ if we identify $A^{i}$, $\psi^{A}$, $\Sigma^{AB}=
\Sigma^{i}(\frac{\sigma^{i}}{2i})^{AB}$ and
$\chi^{A}$ with the anti-self-dual part of the spin connection,
the left-handed gravitino, the chiral two-form $e^{A}_{A^{\prime}}
\wedge e^{BA^{\prime}}$ constructed from
the vierbein $e^{AA^{\prime}}=e^{AA^{\prime}}_{\mu}dx^{\mu}$,
and with the chiral two form $e^{A}_{A^{\prime}}\wedge
\psi^{A^{\prime}}$ constructed from the right-handed
gravitino $\psi^{A^{\prime}}$, respectively\footnote{
Our action is in fact twice the action which is used in
refs.\cite{sano}\cite{shira}}.
As a consequence of this identification
the components $(\Sigma^{i},\chi^{A})$ of the $\CB$
field are subject to the algebraic constraints
\beqy
\Sigma^{ABCD}&\equiv&\Sigma^{(AB}\wedge\Sigma^{CD)}=0,\nonumber \\
\Xi^{ABC}&\equiv&\Sigma^{(AB}\wedge\chi^{C)}.\label{eq:alcon}
\eeqy

In order to obtain the action for Ashtekar's formalism
we first solve the algebraic constraints (\ref{eq:alcon}) for
the time components $(\Sigma^{i}_{ta},\chi^{A}_{ta})$, and
then substitute the result into the canonical action(\ref{eq:caac1}).
General solutions to eq.(\ref{eq:alcon}) are given by
\beqy
\Sigma^{i}_{ta}&=&-\frac{1}{2}\utep_{abc}
(-i\tN\frac{\ep^{ijk}}{2}\tpi^{bj}\tpi^{ck}+2N^{b}\tpi^{ci}),
\nonumber \\
\chi^{A}_{ta}&=&-\utep_{abc}(-i\tN\tpi^{bA}_{B}\tpi^{cB}+
N^{b}\tpi^{cA})+\utep_{abc}\tpi^{bA}_{B}\tpi^{cBC}\tM_{C}.
\eeqy
By substituting this expression into eq.(\ref{eq:caac1}), we find
\beqy
-iI_{Ash}^{N=1}&=&\int dt\int_{\M3}d^{3}x(\tpi^{ai}\dot{A}^{i}_{a}
+2\tpi^{aA}\dot{\psi}_{aA} \nonumber \\
& &+A_{t}^{i}G^{i}-2\psi_{tA}L^{A}
+2\tM_{A}R^{A}+i\tN\CH-2N^{a}\CH_{a}),
\eeqy
with the new constraints
\beqy
R^{A}&=&\frac{1}{2}\utep_{abc}\ep^{ijk}\tpi^{bj}\tpi^{ck}
(\frac{\sigma^{i}}{2i})^{AB}\Phi^{a}_{B},\nonumber \\
\CH&=&\frac{1}{4}\utep_{abc}\ep^{ijk}\tpi^{bj}\tpi^{ck}\Phi^{ai}
-2\utep_{abc}\tpi^{bi}(\frac{\sigma^{i}}{2i})_{AB}\tpi^{cB}\Phi^{aA},
\nonumber \\
\CH_{a}&=&\frac{1}{2}\utep_{abc}\tpi^{bi}\Phi^{ci}+
\utep_{abc}\tpi^{bA}\Phi^{c}_{A}.
\eeqy
Physically, $R^{A}$ generates right-supersymmetry
transformations, $\CH$ generates bubble-time evolutions, and
$\CH_{a}$ generates spatial diffeomorphisms. In passing,
among the $GSU(2)$ gauge transformations, the transformations
generated by $G^{i}$ are reinterpreted as local Lorentz
transformations for left-handed fields and the transformations
generated by $L^{A}$ are regarded as left-supersymmetry
transformations.

Let us now briefly consider the canonical quantization.
Here we also use $\Psi[\CA_{a}]$ as wavefunctions. Because Gauss'
law constraint remains intact, the wavefunctions have to be
invariant under the $GSU(2)$ gauge transformations. When we solve
the new constraints $(R^{A},\CH,\CH_{a})$, we should note that
these constraints are linear combinations of the constraints
$(\Phi^{ai},\Phi^{aA})$ in the BF theory, with the coefficients
being polynomials in the momenta $(\tpi^{ai},\tpi^{aA})$.
As a consequence, if we take the ordering with the momenta to the
left, the solutions (\ref{eq:topo1})(\ref{eq:CS1}) for the $GSU(2)$
BF theory are involved into the solution space for quantum
$N=1$ supergravity in the Ashtekar form. These solutions
are the topological solutions found in
refs.\cite{mats}\cite{sano}\cite{shira}.

Before ending this section we see how the symmetry of the theory
is influenced by the algebraic constraints(\ref{eq:alcon}).
For this purpose we look into the variation of the constraints
$(\Sigma^{ABCD},\Xi^{ABC})$ under the gauge transformations (in a
broader sense). Because these constraints transform covariantly
under the $GSU(2)$ gauge transformations
(see the appendix), the $GSU(2)$
gauge symmetry is preserved even after imposing the algebraic
constraints. However, the Kalb-Ramond symmetry (\ref{eq:KR1}) in
general breaks down because the variation of
$(\Sigma^{ABCD},\Xi^{ABC})$ does not vanish even after imposing all
the constraints. More precisely, by computing the variation using eq.
(\ref{eq:kr1}) and equations of motion which are derived from the
variation of the action(\ref{eq:BFac}) w.r.t. the connection $\CA$,
we find
\beqy
\delta_{\xi}\Sigma^{ABCD}&=&-2D(\phi^{(AB}\wedge\Sigma^{CD)})
+2\{\phi^{(AB}\wedge\chi^{C}+
\Sigma^{(AB}\wedge\eta^{C}\}\wedge\psi^{D)}\nonumber \\
\delta_{\xi}\Xi^{ABC}&=&-D(\phi^{(AB}\wedge\chi^{C)}+
\Sigma^{(AB}\wedge\eta^{C)})-2\lambda g\phi^{(AB}
\wedge\Sigma^{CD)}\psi_{D}. \label{eq:vari1}
\eeqy
In other words, the Kalb-Ramond symmetry survives if the parameter
$\xi$ is such that the variation(\ref{eq:vari1}) vanishes.
A sufficient condition for not violating the Kalb-Ramond symmetry
is provided by
\beq
\phi^{(AB}\wedge\Sigma^{CD)}=
\phi^{(AB}\wedge\chi^{C)}+\Sigma^{(AB}\wedge\eta^{C)}=0.
\label{eq:suffi1}
\eeq
If we assume the vierbein $e^{AA^{\prime}}$ to be nondegenerate,
this equation is completely solved by the superposition of
the diffeomorphisms
\beqy
\phi^{i}_{\mu}=v^{\nu}\Sigma^{i}_{\mu\nu} \nonumber \\
\eta^{A}_{\mu}=v^{\nu}\chi^{A}_{\mu\nu}, \label{eq:diffeo1}
\eeqy
and the right-supersymmetry transformations
\beqy
\phi^{i}&=&0,\nonumber \\
\Sigma^{(AB}\wedge\eta^{C)}&=&0. \label{eq:RSUSY1}
\eeqy
Thus we have seen explicitly that the imposition of the algebraic
constraints(\ref{eq:alcon}) breaks the Kalb-Ramond symmetry down
to the symmetry under the diffeomorphisms and the
right-supersymmetry transformations.

In the Lagrangian formalism, we impose the algebraic constraints
by introducing the linear terms in the auxiliary fields
$(\Psi_{ABCD}=\Psi_{(ABCD)},\kappa_{ABC}=\kappa_{(ABC)})$:
\beq
-iI^{N=1}_{aux.}=\int(-\Psi_{ABCD}\Sigma^{AB}\wedge\Sigma^{CD}
-2\kappa_{ABC}\Sigma^{AB}\wedge\chi^{C}).\label{eq:aux1}
\eeq
The transformation properties of the fields are somewhat modified,
while the essential features remain valid.
This is explained in the Appendix.


\section{$G^{2}SU(2)$ BF theory and Ashtekar's formulation
for $N=2$ supergravity}

In this section we demonstrate
that $N=2$ supergravity can be cast into
the form of the \lq\lq constrained" BF theory with the gauge group
being an appropriate graded version of $SU(2)$.
Except a few subtleties, the argument goes in almost the same
manner as in the previous two cases. So we briefly explain the
overview focusing on the subtleties.

The relevant graded Lie algebra is provided by
\beqy
[J_{i},J_{j}]&=& \ep_{ijk}J_{k},\quad
[J_{i},J_{A}^{(\alpha)}]=(\frac{\sigma^{i}}{2i})_{A}\UI{B}
J_{B}^{(\alpha)},\quad [J_{i},J]=0,
\nonumber \\
{}[J,J_{A}^{(\alpha)}]&=& g(\tau^{3})^{\alpha}\LI{\beta}
J_{A}^{(\beta)},\quad[J,J]=0, \nonumber \\
\{J_{A}^{(\alpha)},J_{B}^{(\beta)}\}&=&
-\ep^{\alpha\beta}\ep_{AB}J+4g(\tau^{3})^{\alpha\beta}
(\frac{\sigma^{i}}{2i})_{AB}J_{i},\label{eq:G2SU2}
\eeqy
where $\alpha,\beta,\cdots$ denotes the spinor indices for
the internal $SU(2)$ symmetry existing in the $N=2$ supergravity
with a vanishing cosmological constant $\Lambda\equiv -6g^{2}=0$.
$\tau^{3}$ is the third
component of the Pauli matrices:
\beq
(\tau^{3})^{\alpha}\LI{\beta}=(\tau^{3})_{\beta}\UI{\alpha}
\equiv(\tau^{3})^{\alpha}_{\beta}=\left(\begin{array}{cc}
1&0 \\ 0&-1 \end{array}\right).
\eeq
In this paper we will tentatively refer to this graded algebra
(\ref{eq:G2SU2}) as $G^{2}SU(2)$.

Let us investigate the $G^{2}SU(2)$ BF theory. The action is
\beq
-iI^{N=2}_{BF}=\int\Str(\CB\wedge\CF+g^{2}\CB\wedge\CB),
\label{eq:BFAC}
\eeq
where $\CB=\Sigma^{i}J_{i}-\frac{1}{2g}(\tau^{3})^{\beta}_{\alpha}
\chi^{A}_{\beta}J_{A}^{(\alpha)}+\frac{1}{4g^{2}}BJ$ is a
$G^{2}SU(2)$-valued two-form, $\CF=\CA+\CA\wedge\CA$ is the
curvature two-form of a $G^{2}SU(2)$ connection $\CA=A^{i}J_{i}+
\psi^{A}_{\alpha}J^{(\alpha)}_{A}+AJ$. $\Str$ used here
is the unique $G^{2}SU(2)$-invariant bilinear two form:
\beqy
\Str(J_{i}J_{j})&\!\!\!=\!\!\!&\delta_{ij},\quad
\Str(J_{A}^{(\alpha)}J_{B}^{(\beta)})=4g\ep_{AB}
(\tau^{3})^{\alpha\beta},\quad \Str(JJ)=4g^{2}, \nonumber \\
\Str(J_{i}J_{A}^{(\alpha)})&\!\!\!=\!\!\!&\Str(J_{i}J)=
\Str(J_{A}^{(\alpha)}J)=0.
\eeqy
We can now rewrite the action (\ref{eq:BFAC}) in terms of the
component fields
\beq
-iI^{N=2}_{BF}=\int\left(\begin{array}{l}
\Sigma^{i}\wedge(F^{i}+2g\psi^{A\alpha}\wedge\psi_{\beta}^{B}
(\tau^{3})_{\alpha}^{\beta}(\frac{\sigma^{i}}{2i})_{AB}) \\
\qquad+2\chi_{\alpha}^{A}\wedge(D\psi_{A}^{\alpha}-g
(\tau^{3})^{\alpha}_{\beta}A\wedge\psi_{A}^{\beta})+B\wedge\hat{F}\\
\qquad\qquad
+g^{2}\Sigma^{i}\wedge\Sigma^{i}-g(\tau^{3})_{\alpha}^{\beta}
\chi^{\alpha}_{A}\wedge\chi^{A}_{\beta}+\frac{1}{4}B\wedge B
\end{array}\right),\label{eq:BF2}
\eeq
where we have set $\hat{F}=dA-\frac{1}{2}\psi_{\alpha}^{A}\wedge
\psi^{\alpha}_{A}$. This action
obviously enjoys the symmetry under the $G^{2}SU(2)$ gauge
transformations
\beqy
\delta_{\rho}\CA&=&-\CD\rho\equiv-d\rho-[\CA,\rho]\nonumber \\
\delta_{\rho}\CB&=&[\rho,\CB]\label{eq:GAUGE2}
\eeqy
with $\rho=\theta^{i}J_{i}+\ep_{\alpha}^{A}J_{A}^{(\alpha)}+\lambda
J$ being a $G^{2}SU(2)$-valued scalar, and the Kalb-Ramond symmetry
\beqy
\delta_{\xi}\CA&=&2g^{2}\xi \nonumber \\
\delta_{\xi}\CB&=&-\CD\xi\equiv -d\xi-\CA\wedge\xi-\xi\wedge\CA
\label{eq:KR2}
\eeqy
with $\xi=\phi^{i}J^{i}-\frac{1}{2g}(\tau^{3})^{\beta}_{\alpha}
\eta^{A}_{\beta}J^{(\alpha)}_{A}+\frac{1}{4g^{2}}\kappa J$ being
a $G^{2}SU(2)$-valued one-form. These transformations written
in terms of component fields are as follows. The $G^{2}SU(2)$
gauge transformations are
\beqy
\delta_{\rho}A^{i}&=&-D\theta^{i}+4g(\frac{\sigma^{i}}{2i})_{AB}
(\tau^{3})^{\alpha\beta}\psi_{\alpha}^{A}\ep_{\beta}^{B}\nonumber \\
\delta_{\rho}\psi_{\alpha}^{A}&=&\theta^{i}(\frac{\sigma^{i}}{2i})
^{A}\LI{B}\psi^{B}_{\alpha}-D\ep_{\alpha}^{A}-
g(\tau^{3})_{\alpha}^{\beta}\ep_{\beta}^{A}A+g\lambda(\tau^{3})
_{\alpha}^{\beta}\psi_{\beta}^{A}\nonumber \\
\delta_{\rho}A&=&-d\lambda-\ep_{\alpha}^{A}\psi_{A}^{\alpha}
\nonumber \\
\delta_{\rho}\Sigma^{i}&=&\ep^{ijk}\theta^{j}\Sigma^{k}+
2\ep^{\alpha}_{B}(\frac{\sigma^{i}}{2i})^{B}\LI{C}\chi_{\alpha}^{C}
\nonumber \\
\delta_{\rho}\chi_{\alpha}^{A}&=&
\theta^{i}(\frac{\sigma^{i}}{2i})^{A}\LI{B}\chi_{\alpha}^{B}+
2g(\tau^{3})^{\beta}_{\alpha}
\ep_{\beta}^{B}(\frac{\sigma^{i}}{2i})^{A}\LI{B}\Sigma^{i}
+\frac{1}{2}\ep_{\alpha}^{A}B+g\lambda(\tau^{3})_{\alpha}^{\beta}
\chi_{\beta}^{A}\nonumber \\
\delta_{\rho}B&=&2g\ep_{A}^{\alpha}(\tau^{3})_{\alpha}^{\beta}
\chi_{\beta}^{A}.\label{eq:gauge2}
\eeqy
In the $N=2$ supergravity, transformations generated by $\theta^{i}$,
by $\ep_{\alpha}^{A}$ and by  $\lambda$ are respectively interpreted
as local Lorentz transformations, left-SUSY transformations and
$U(1)$ gauge transformations. The Kalb-Ramons transformations
for the component fields are given by
\beqy
\delta_{\xi}A^{i}&=&2g^{2}\phi^{i}\nonumber \\
\delta_{\xi}\psi_{\alpha}^{A}&=&
-g(\tau^{3})_{\alpha}^{\beta}\eta_{\beta}^{A} \nonumber \\
\delta_{\xi}A&=&\frac{1}{2}\kappa \nonumber \\
\delta_{\xi}\Sigma^{i}&=&-D\phi^{i}-2(\frac{\sigma^{i}}{2i})_{AB}
\psi_{\alpha}^{A}\wedge\eta^{C\alpha}\nonumber \\
\delta_{\xi}\chi_{\alpha}^{A}&=&2g(\tau^{3})_{\alpha}^{\beta}
(\frac{\sigma^{i}}{2i})^{A}\LI{B}\phi^{i}\wedge\psi_{\beta}^{B}
-D\eta_{\alpha}^{A}-g(\tau^{3})_{\alpha}^{\beta}A\wedge
\eta_{\beta}^{A}-\frac{1}{2}\psi_{\alpha}^{A}\wedge\kappa\nonumber \\
\delta_{\xi}B&=&-2g(\tau^{3})_{\alpha}^{\beta}\psi^{\alpha}_{A}
\wedge\eta^{A}_{\beta}-d\kappa.\label{eq:kr2}
\eeqy
As we will see shortly, these transformations are closely related
to the diffeomorphisms and the right-SUSY transformations in the
$N=2$ supergravity.

Let us now briefly look into the canonical quantization.
In the canonical formalism, action (\ref{eq:BFAC}) is rewritten as
follows
\beq
-iI^{N=2}_{BF}=\int dt\int_{\M3}d^{3}x\Str[\tPi^{a}\dot{\CA}_{a}
+\CA_{t}{\bf G}+\CB_{ta}{\bf\Phi}^{a}], \label{eq:CAAC2}
\eeq
where we have set $\tPi^{a}=\frac{1}{2}
\otep^{abc}\CB_{bc}=\tpi^{ai}J_{i}
-\frac{1}{2g}(\tau^{3})_{\alpha}^{\beta}\tpi^{Aa}_{\beta}
J_{A}^{(\alpha)}+\frac{1}{4g^{2}}\tpi^{a}J$. In terms of
the component fields this canonical action becomes
\beq
-iI^{N=2}_{BF}=\int dt\int_{\M3}d^{3}x\left(\begin{array}{l}
\tpi^{ai}\dot{A}_{a}^{i}+2\tpi^{aA}_{\alpha}\dot{\psi}^{\alpha}_{Aa}
+\tpi^{a}\dot{A}_{a} \\ \qquad
+A_{t}^{i}G^{i}-2\psi^{\alpha}_{At}L^{A}_{\alpha}+A_{t}G \\
\qquad\qquad +\Sigma^{i}_{ta}\Phi^{ai}+2\chi^{A}_{\alpha ta}
\Phi^{a\alpha}_{A}+B_{ta}\Phi^{a}
\end{array}\right).\label{eq:caac2}
\eeq
As in the previous cases this system has two types of first class
constraints. Gauss' law constraint
\beqy
{\bf G}&=&\CD_{a}\tPi^{a} \nonumber \\
&=&G^{i}J_{i}-\frac{1}{2g}(\tau^{3})_{\alpha}^{\beta}
L_{\beta}^{A}J_{A}^{(\alpha)}+\frac{1}{4g^{2}}GJ \label{eq:GAUSS2}
\eeqy
generates the $G^{2}SU(2)$ transformations(\ref{eq:gauge2}).
And the remaining constraint
\beqy
{\bf\Phi}^{a}&=&\frac{1}{2}\otep^{abc}\CF_{bc}+2g^{2}\tPi^{a}
\nonumber \\
&=&\Phi^{ai}J_{i}+\Phi^{aA}_{\alpha}J_{A}^{\alpha}+\Phi^{a}J
\eeqy
generates the Kalb-Ramond symmetry(\ref{eq:kr2}). The explicit form
of Gauss' law constraint can be seen in refs.\cite{kuni}\cite{sano}.
Expressions for the remaining constraints are also seen implicitly
in these references.

Canonical quantization of this theory can be handled in the analogous
way to the $GSU(2)$ case (except for $g=0$).
We will use as wavefunctions the functionals $\Psi[\CA_{a}]$ of
the $G^{2}SU(2)$ connection $\CA_{a}dx^{a}$.
Gauss' law constraint tells us that $\Psi[\CA_{a}]$ should be
invariant under (the identity-connected component of) the
$G^{2}SU(2)$ gauge transformations. For $g\neq 0$, the remaining
constraint can be solved similarly to the $GSU(2)$ case.
In this case we have the unique solution
\beqy
\Psi[\CA_{a}]&=&e^{\frac{1}{4g^{2}}W_{CS}^{N=2}},\nonumber \\
W_{CS}^{N=2}&=&\int_{\M3}\Str(\CA d\CA+\frac{2}{3}\CA\wedge\CA
\wedge\CA) \nonumber \\
&=&\int_{\M3}[A^{i}dA^{i}+\frac{1}{3}\ep^{ijk}A^{i}\wedge A^{j}
\wedge A^{k}\nonumber \\
& &\qquad-4g(\tau^{3})^{\alpha\beta}\psi_{\alpha A}\wedge
(D\psi_{\beta}^{A}+g(\tau^{3})^{\gamma}_{\beta}A\wedge
\psi_{\gamma}^{A})+4g^{2}AdA].\label{eq:CS2}
\eeqy
This coincides with the $N=2$ super-extended version of the
Chern-Simons solution found in ref.\cite{sano}.

For $g=0$, a special consideration is needed, because in the present
case a remnant of the \lq\lq cosmological term"
$g^{2}\CB\wedge\CB$ exists even in the
limit $g\rightarrow 0$. Particularly, the reduced phase space
loses the cotangent bundle structure unlike in the case of
$SU(2)$ or $GSU(2)$. We can nevertheless construct solutions
to the quantum constraints, at least formally. As in the $g\neq 0$
case Gauss' law constraint merely requires the wavefunctions to be
$G^{2}SU(2)$ gauge invariant. The remaining constraints in the $g=0$
case is written as
\beqy
\hat{\Phi}^{ai}\cdot\Psi[\CA]&=&\frac{1}{2}\otep^{abc}F^{i}_{bc}
\cdot\Psi[\CA]=0, \nonumber \\
\hat{\Phi}_{A}^{a\alpha}\cdot\Psi[\CA]&=&\otep^{abc}
D_{b}\psi^{\alpha}_{Ac}\cdot\Psi[\CA]=0, \nonumber \\
\hat{\Phi}^{a}\cdot\Psi[\CA]&=&\left[
\frac{1}{2}\otep^{abc}(F_{bc}-\psi^{A}_{\alpha b}\psi^{\alpha}_{Ac})
-\frac{1}{2}\frac{\delta}{\delta A_{a}}\right]\Psi[\CA]=0,
\label{eq:const2}
\eeqy
where $F_{bc}=2\partial_{[b}A_{c]}$ is the field strength of the
$U(1)$ connection $A$.
Formal solutions to these equations are given by
\beq
\Psi[\CA]=e^{W_{U(1)}}F[A^{i},\psi_{A}^{\alpha}]\prod_{x\in\M3}
\left(\prod_{a,i}\delta(\otep^{abc}F^{i}_{bc}(x))\prod_{a,A,\alpha}
\delta(\otep^{abc}D_{b}\psi_{Ac}^{\alpha}(x))\right),\label{eq:fsol}
\eeq
where $F[A^{i},\psi_{A}^{\alpha}]$ is a $G^{2}SU(2)$ gauge invariant
function of $(A^{i}_{a},\psi_{Aa}^{\alpha})$ and
$$
W_{U(1)}\equiv\int_{\M3}(AdA-A\wedge\psi_{\alpha}^{A}\wedge
\psi^{\alpha}_{A}).
$$
As it is, however, eq.(\ref{eq:fsol}) is not
$G^{2}SU(2)$ gauge invariant. There is no problem in the
delta function part because the curvatures $(F^{i},
D\psi^{\alpha}_{A})$ transform covariantly under the $G^{2}SU(2)$
gauge transformations and because their gauge transformations do not
involve the $U(1)$ part $\hat{F}=dA-\frac{1}{2}\psi_{\alpha}^{A}
\wedge\psi^{\alpha}_{A}$. The functional $W_{U(1)}$ is, however,
not invariant under the left-SUSY transformations. After a somewhat
lengthy calculation, we see that $W_{U(1)}$ transforms as\footnote{
We have assumed that $F^{i}_{bc}=D_{[b}\psi^{\alpha}_{c]A}=0$ hold.}
\beqy
&&e^{-i\hat{L}(\ep)}W_{U(1)}e^{i\hat{L}(\ep)}-W_{U(1)} \nonumber \\
&&=\int_{\M3}[\ep_{\alpha}^{A}\psi_{A}^{\alpha}\wedge\psi_{\beta}^{B}
\wedge\psi^{\beta}_{B}-\frac{1}{2}\ep_{A}^{\alpha}D\ep_{\alpha}^{A}
\wedge\psi_{\beta}^{B}\wedge\psi_{B}^{\beta}-\ep_{\alpha}^{A}
\psi_{A}^{\alpha}\wedge d(\ep_{\beta}^{B}\psi_{B}^{\beta})
\nonumber \\
&&\qquad\qquad\qquad
+D\ep_{\alpha}^{A}\wedge D\ep^{\alpha}_{A}\wedge\ep_{\beta}^{B}
\psi_{B}^{\beta}-\frac{1}{4}\ep_{A}^{\alpha}D\ep_{\alpha}^{A}
\wedge d(\ep_{B}^{\beta}D\ep_{\beta}^{B})]
\eeqy
where $\hat{L}(\ep)\equiv-2i\int_{\M3}d^{3}x\ep^{A}_{\alpha}
\hat{L}_{A}^{\alpha}$ is the generator of the left-SUSY
transformations. We should note that the $U(1)$ connection $A$
does not appear anywhere in the r.h.s of the above expression.
The wavefunction (\ref{eq:fsol}) with $W_{U(1)}$ replaced by
$e^{-i\hat{L}(\ep)}W_{U(1)}e^{i\hat{L}(\ep)}$ therefore
remains to be the solution of eq.(\ref{eq:const2}). Now we can
give formal solutions to all the constraint equations in the
$g=0$ case:
\beqy
\Psi[\CA]&=&F[A^{i},\psi_{A}^{\alpha}]\prod_{x\in\M3}
\left(\prod_{a,i}\delta(\otep^{abc}F^{i}_{bc}(x))\prod_{a,A,\alpha}
\delta(\otep^{abc}D_{b}\psi_{Ac}^{\alpha}(x))\right) \nonumber \\
& &\times\int[d\ep_{\alpha}^{A}]\exp(e^{-i\hat{L}(\ep)}
W_{U(1)}e^{i\hat{L}(\ep)}),\label{eq:FSOL}
\eeqy
where $[d\ep_{\alpha}^{A}]$ denotes an $SU(2)$ invariant measure.

In passing, $F[A^{i},\psi_{A}^{\alpha}]$ can be interpreted as
the gauge invariant functional of the \lq\lq truncated" connection
$\widehat{\CA}\equiv A^{i}J_{i}+\psi^{A}_{\alpha}
\widehat{J}_{A}^{(\alpha)}$, where
$(J_{i},\widehat{J}_{A}^{(\alpha)})$
are the generators of the following truncated algebra
\beq
[J_{i},J_{j}]=\ep_{ijk}J_{k},\quad
[J_{i},\widehat{J}_{A}^{(\alpha)}]=(\frac{\sigma^{i}}{2i})_{A}\UI{B}
\widehat{J}_{B}^{(\alpha)},\quad
\{\widehat{J}_{A}^{(\alpha)},\widehat{J}_{B}^{(\beta)}\}=0.
\eeq
This is possible
because the $U(1)$ part $J$ in the $G^{2}SU(2)$ algebra
(\ref{eq:G2SU2}) with $g=0$ almost decouples from the
rest generators $(J_{i},J_{A}^{(\alpha)})$.

The relation to $N=2$ supergravity is not so simple. This is because
the chiral action of $N=2$ supergravity\cite{kuni}\cite{sano}
\beqy
-iI^{N=2}_{Ash}&\!\!\!=\!\!\!&\int\left(\begin{array}{l}
\Sigma^{i}\wedge(F^{i}+2g(\tau^{3})_{\alpha}^{\beta}
(\frac{\sigma^{i}}{2i})_{AB}\psi^{\alpha A}\wedge\psi^{B}_{\beta})
+2\chi_{\alpha}^{A}\wedge(D\psi_{A}^{\alpha}-g
(\tau^{3})^{\alpha}_{\beta}A\wedge\psi_{A}^{\beta})\\
+g^{2}\Sigma^{i}\wedge\Sigma^{i}-g(\tau^{3})_{\alpha}^{\beta}
\chi_{A}^{\alpha}\wedge\chi_{\beta}^{A}-\Psi_{ABCD}\Sigma^{AB}\wedge
\Sigma^{CD}
-2\kappa_{ABC}^{\alpha}\Sigma^{AB}\wedge\chi_{\alpha}^{C}\\
\qquad -\hat{F}\wedge\hat{F}+
\varphi^{i}\hat{F}\wedge\Sigma^{i}-\frac{1}{4}
\varphi^{i}\varphi^{j}\Sigma^{i}\wedge\Sigma^{j}+\varphi^{i}
(\frac{\sigma^{i}}{2i})_{AB}\chi_{\alpha}^{A}\wedge\chi^{B\alpha}
\end{array}\right)\nonumber \\
& &\quad \label{eq:ASH2}
\eeqy
involves the terms which are (at most) quadratic in the auxiliary
field $\varphi^{i}$:
\beq
-iL_{U(1)}\equiv
-\hat{F}\wedge\hat{F}+\varphi^{i}\hat{F}\wedge\Sigma^{i}-\frac{1}{4}
\varphi^{i}\varphi^{j}\Sigma^{i}\wedge\Sigma^{j}+\varphi^{i}
(\frac{\sigma^{i}}{2i})_{AB}\chi_{\alpha}^{A}\wedge\chi^{B\alpha}.
\label{eq:U1ac}
\eeq
First we translate this quadratic part $-iL_{U(1)}$ into the terms
which are at most linear in auxiliary fields as follows.
We know how to deal with auxiliary fields which appear in the
action at most quadratically: we have only to solve equations
of motion obtained from the variation w.r.t the auxiliary
fields and to substitute the result into the action.
In the present case the desired equations of motion are
\beq
\Sigma^{i}\wedge B=2(\frac{\sigma^{i}}{2i})_{AB}
\chi_{\alpha}^{A}\wedge\chi^{B\alpha},\label{eq:eom}
\eeq
where we have set $B=\varphi^{i}\Sigma^{i}-2\hat{F}$. Using this,
eq.(\ref{eq:U1ac}) is rewritten as
$$
-iL_{U(1)}=-\frac{1}{4}B\wedge B+(B+2\hat{F})\wedge\frac{1}{2}B.
$$
Arranging this expression neatly and taking account of the
algebraic constraint (\ref{eq:eom}), the quadratic part
(\ref{eq:U1ac}) turns out to be equivalent to the following
expression which is at most linear in the new auxiliary field
$\varphi^{\prime}_{AB}$ \footnote{$\varphi^{\prime}_{AB}$ can be
identified with $\varphi_{AB}=\varphi^{i}
(\frac{\sigma^{i}}{2i})_{AB}$ if $B$ is integrated out.}
\beq
-iL_{U(1)}=B\wedge\hat{F}+\frac{1}{4}B\wedge B-\varphi^{\prime}_{AB}
(\Sigma^{AB}\wedge B-\chi_{\alpha}^{A}\wedge\chi^{B\alpha}).
\eeq
Substituting this into the chiral $N=2$ action (\ref{eq:ASH2}) and
comparing the result with the $G^{2}SU(2)$ BF action (\ref{eq:BF2}),
we find
\beqy
-iI^{N=2}_{Ash}&=&-iI^{N=2}_{BF}-iI^{N=2}_{aux.}\nonumber \\
-iI^{N=2}_{aux.}&=&\int[-\Psi_{ABCD}\Sigma^{AB}\wedge\Sigma^{CD}
-2\kappa^{\alpha}_{ABC}\Sigma^{AB}\wedge\chi_{\alpha}^{C}
\nonumber \\
& &\qquad-\varphi^{\prime}_{AB}(\Sigma^{AB}\wedge B-
\chi_{\alpha}^{A}\wedge\chi^{B\alpha})].
\eeqy
Thus we have established the relation between $N=2$ supergravity
and the $G^{2}SU(2)$ BF theory. Namely, $N=2$ supergravity in
Ashtekar's form is regarded as the $G^{2}SU(2)$ BF theory
(\ref{eq:BFAC}), with the $\CB$ fields being subject to
the algebraic constraints
\beqy
\Sigma^{ABCD}&\equiv&\Sigma^{(AB}\wedge\Sigma^{CD)}=0\nonumber \\
\Xi^{ABC}_{\alpha}&\equiv&
\Sigma^{(AB}\wedge\chi_{\alpha}^{C)}=0 \nonumber \\
{\bf B}^{AB}&\equiv&\Sigma^{AB}\wedge B-\chi_{\alpha}^{A}\wedge
\chi^{B\alpha}=0.\label{eq:alcon2}
\eeqy

As in the $N=1$ case, Ashtekar's formalism for $N=2$ supergravity
is derived by solving these algebraic constraints for
the time components $(\Sigma^{i}_{ta},\chi^{A}_{\alpha ta},B_{ta})$
and by substituting the solution into the canonical BF action
(\ref{eq:caac2}). Gauss' law constraint is inherited as it is
from the BF theory. In addition we have three types of constraints:
the Hamiltonian constraint $\CH$, the diffeomorphism
constraint $\CH_{a}$ and the constraint $R^{A}_{\alpha}$ which
generates right-SUSY transformations. These are given by
linear combinations of the constraints $(\Phi^{ai},
\Phi^{a\alpha}_{A},\Phi^{a})$ in the $G^{2}SU(2)$ BF theory:
\beq
\CH^{{\cal I}}=C^{{\cal I}}_{ai}(\tPi^{a})\Phi^{ai}+
C^{{\cal I}A}_{a\alpha}(\tPi^{a})\Phi^{a\alpha}_{A}+
C^{{\cal I}}_{a}(\tPi^{a})\Phi^{a},
\eeq
where we have set
$(\CH^{{\cal I}})\equiv(\CH,\CH_{a},R_{\alpha}^{A})$.
The crucial thing is that the coefficients depend only on
the momenta $\tPi^{a}$ and not on the connections $\CA_{a}$.
The solutions (\ref{eq:CS2})(\ref{eq:FSOL}) to the quantum
$G^{2}SU(2)$ BF
theory are thus included in the solution space of canonically
quantized $N=2$ supergravity, provided that we take the ordering
with the momenta to the left.

Similarly to the $N=1$ case the Kalb-Ramond symmetry (\ref{eq:kr2})
in general breaks down owing to the algebraic constraint
(\ref{eq:alcon2}). By the argument parallel to the previous section
we can find a sufficient condition for the Kalb-Ramond symmetry to
preserve the algebraic constraints:
\beqy
\phi^{(AB}\wedge\Sigma^{BC)}&=&0,\nonumber \\
\phi^{(AB}\wedge\chi_{\alpha}^{C)}+
\Sigma^{(AB}\wedge\eta^{C)}_{\alpha}&=&0, \nonumber \\
\phi^{AB}\wedge B+\Sigma^{AB}\wedge\kappa&=&2\eta_{\alpha}^{(A}
\wedge\chi^{B)\alpha}.
\eeqy
Assuming that the vierbein $e^{AA^{\prime}}$ to be nondegenerate,
these equations are completely solved by the superposition of
the diffeomorphisms
\beq
\phi^{i}_{\mu}=v^{\nu}\Sigma^{i}_{\mu\nu},\quad
\eta^{A}_{\alpha\mu}=v^{\nu}\chi^{A}_{\alpha\mu\nu},\quad
\kappa_{\mu}=v^{\nu}B_{\mu\nu},
\eeq
and the right-SUSY transformations
\beq
\phi^{i}=0,\quad
\Sigma^{(AB}\wedge\eta^{C)}_{\alpha}=0,\quad
\Sigma^{AB}\wedge\kappa=2\eta_{\alpha}^{(A}\wedge\chi^{B)\alpha}.
\eeq


\section{Discussion}

In this paper we have shown explicitly that
$N=1$ and $N=2$ supergravities in Ashtekar's form can be
cast into the form of $BF$ theories with the $B$ fields
subject to the algebraic constraints. Once we have established
these relations it is expected that considerable progress
will be made on the canonical quantum gravity both technically
and conceptually.

For example,
we may use the technic developed in the BF theory\cite{horo}
at least when we investigate the topological sector of the canonical
quantum gravity. With regard to pure gravity some works of this kind
can be seen in refs. \cite{ueno}\cite{ueno2}\cite{cotta}. The results
in this paper suggest that we can exploit similar methods also
for studying $N=1,2$ supergravities.
Because the BF theory resembles the Chern-Simons
gauge theory\cite{witt}, the methods for studying (2+1)-dimensional
Einstein gravity in the Chern-Simons form\cite{town}\cite{witt2}
may be applied.
It is of particular interest to investigate the physical significance
of the topological solutions. While geometrical interpretation of
the Chern-Simons solutions is studied considerably well
\cite{kodama}\cite{shira}\cite{soo}, we do not know any works
on the geometrical interpretation of the topological solutions
in the case where the cosmological constant vanishes.
Naively considering these solutions correspond to the flat spaces
because they have support only on the flat anti-self-dual
connections and because the imposition of the reality condition
indicates that the self-dual connection should also be flat.
The problem is how the moduli of the connections are related to
the spacetime structure. We anticipate that they are intimately
related with the geometric structures as in (2+1)-dimensions
\cite{carl}. This is currently under investigation\cite{ezawa2}.

Recently an attempt appeared to extend the loop representation
\cite{smol} to $N=1$ supergravity \cite{gam}. As we have
shown that $N=2$ supergravity is described by the $G^{2}SU(2)$
connection, loop representation may be extended also to $N=2$
supergravity.

There are several attempts to interpret Einstein gravity as an
\lq\lq unbroken phase" of some topological field theories
\cite{medi}\cite{suga}. We may extend these ideas to
supergravities. Probably this deserves studying because
the existence of the supersymmetry is believed by
many people and thus supergravities seem to be more realistic than
pure gravity.

$N=2$ supergravity is of its own interest because the twisted
version of $N=2$ supergravity gives rise to a topological gravity
\cite{anse}. Ashtekar's formalism can be applied also to
this twisted $N=2$ supergravity \cite{paul}. To see whether twisted
$N=2$ supergravity is related to a BF theory or not is left to the
future investigation.

\vspace{0.2in}

\noindent Acknowledgments

I would like to thank Prof. K. Kikkawa,
Prof. H. Itoyama and H. Kunitomo
for useful discussions and careful readings of the manuscript.


\appendix

\catcode`\@=11
\def\theequation{\Alph{section}.\arabic{equation}}

\section{Appendix}

In this appendix we look into the symmetry of the $N=1$ Ashtekar's
formalism in the Lagrangian form. The relevant action is
\beq
-iI^{N=1}_{Ash}=-iI^{N=1}_{BF}-iI^{N=1}_{aux.},\label{eq:action1}
\eeq
where $-iI^{N=1}_{BF}$ is the $GSU(2)$ BF action(\ref{eq:BFac})
and $-iI^{N=1}_{aux.}$ is the linear terms (\ref{eq:aux1}) in the
auxiliary fields $(\Psi_{ABCD},\kappa_{ABC})$. When we discuss
the symmetry of the action, we cannot use equations of motion.
This is because the equations of motion are nothing but the
condition for the action to be stationary under any variations
of the fields. Thus a careful consideration is necessary.

In order to $I^{N=1}_{Ash}$ make invariant under the $GSU(2)$
gauge transformations, we have only to consider
$(\Psi_{ABCD},\kappa_{ABC})$ to be covariant under these
transformations. For $SU(2)$ transformations, it is obvious
that $(\Psi_{ABCD},\kappa_{ABC})$ should transform as is
suggested by their
spinor indices. Because the algebraic constraints
transform under the left-SUSY transformations as
\beqy
\delta_{\ep}\Sigma^{ABCD}&=&-2\ep^{(A}\Xi^{BCD)} \nonumber \\
\delta_{\ep}\Xi^{ABC}&=&-\lambda g\Sigma^{ABCD}\ep_{D},
\eeqy
the auxiliary fields are required to transform  as follows
\beqy
\delta_{\ep}\Psi_{ABCD}&=&2\lambda g\kappa_{(ABC}\ep_{D)}
\nonumber \\
\delta_{\ep}\kappa_{ABC}&=&\Psi_{ABCD}\ep^{D}.
\eeqy

Next we consider the Kalb-Ramond symmetry. The transformations
of the algebraic constraints under off-shell are
\beqy
\delta_{\xi}\Sigma^{ABCD}&=&-2\phi^{(AB}\wedge\{
D\Sigma^{CD)}-\psi^{C}\wedge\chi^{D)}\}+(\mbox{terms appeared in
eq.(\ref{eq:vari1})}),\nonumber \\
\delta_{\xi}\Xi^{ABC}&=&-\phi^{(AB}\wedge\{
D\chi^{C)}-\lambda g\Sigma^{C)D}\wedge\psi_{D}\}+
\{D\Sigma^{(AB}-\psi^{(A}\wedge\chi^{B}\}\wedge\eta^{C)}\nonumber \\
& &+(\mbox{terms appeared in eq.(\ref{eq:vari1})}).
\eeqy
We should be aware that the expressions in the braces are
the equations of motion obtained from the variation of the
action (\ref{eq:action1}) w.r.t the
connection $\CA$. This implied that, under the condition
(\ref{eq:suffi1}), we can render the action $-iI^{N=1}_{Ash}$
invariant by adding some extra terms to the transformation
of the connection. Adding these extra terms to the original
transformations(\ref{eq:kr1}), we find the total transformation
of the connection
\beqy
\delta_{\xi}A^{i}&=&-\frac{g^{2}}{3}\phi^{i}-2(
\frac{\sigma^{i}}{2i})^{AB}(\Psi_{ABCD}\phi^{CD}
+\kappa_{ABC}\eta^{C}), \nonumber \\
\delta_{\xi}\psi^{A}&=&\frac{g}{3\lambda}\eta^{A}
+\kappa^{ABC}\phi_{BC}.\label{eq:modikr}
\eeqy
If we set $\phi^{i}=0$, this exactly coincides with the right-SUSY
transformation in refs.\cite{sano}\cite{shira}.
We can also show that, similarly to the cases of BF theories,
the transformation (\ref{eq:modikr}) with the parameter
$\xi_{\mu}=v^{\nu}\CB_{\mu\nu}$ yields the diffeomorphism
generated by $v^{\mu}\frac{\partial}{\partial x^{\mu}}$
under on-shell.


\end{document}